\documentclass[a4paper,11pt]{article}

\usepackage{amsmath,amssymb,mathtools}
\usepackage{xcolor}
\usepackage{color}
\usepackage{graphicx}
\usepackage{subfigure}
\usepackage{cite}
\usepackage[colorlinks=true,linkcolor=red,citecolor=blue,urlcolor=blue,bookmarks]{hyperref}
\usepackage{multirow,makecell}
\usepackage{textcomp}
\usepackage{wasysym}
\usepackage{ulem}

\usepackage{pifont}

\usepackage{verbatim}

\usepackage[utf8]{inputenc}
\usepackage[T1]{fontenc}

\usepackage{array}
\usepackage{tikz}
\usetikzlibrary{arrows,decorations.pathreplacing,positioning,calc}

\newcommand{\email}[1]{\href{mailto:#1}{\textcolor{black}{\nolinkurl{#1}}}}

\usepackage[text={17cm,25cm},centering]{geometry} %% thanks to Chao Wu
\numberwithin{equation}{section}

\def \be {\begin{equation}}
\def \ee {\end{equation}}
\def \ba {\begin{array}}
\def \ea {\end{array}}
\def \bea {\begin{eqnarray}}
\def \eea {\end{eqnarray}}
\def \nn {\nonumber}

\def \a {\alpha}

\def \D {\Delta}
\def \e {\epsilon}
\def \ve {\varepsilon}

\def \m {\mu}
\def \n {\nu}

\def \s {\sigma}

\def \r {\rho}

\def \th {\theta}

\def \t {\tau}

\def \cO {\mathcal O}
\def \cP {\mathcal P}
\def \cQ {\mathcal Q}

\def \rR {\mathrm R}

\def \f {\frac}

\def \lt {\left}
\def \rt {\right}

\def \lag {\langle}
\def \rag {\rangle}

\def \dd {\mathrm{d}}
\def \ep {\mathrm{e}}
\def \ii {\mathrm{i}}

\def \tr {\textrm{tr}}

\def \and {{~\textrm{and}~}}

\def \NS {{\textrm{NS}}}

\def \I {{\textrm{I}}}
\def \II {{\textrm{II}}}

\def \NS {{\textrm{NS}}}

\def \I {{\textrm{I}}}
\def \II {{\textrm{II}}}
\def \III {{\textrm{III}}}
\def \IV {{\textrm{IV}}}
\def \V {{\textrm{V}}}
\def \VI {{\textrm{VI}}}
\def \VII {{\textrm{VII}}}
\def \VIII {{\textrm{VIII}}}
\def \IX {{\textrm{IX}}}

\begin{document}

\title{\textbf{ A new product entropy }}

\author{
Jiaju Zhang$^{a}$\footnote{\email{jiajuzhang@tju.edu.cn}}~,
Zhuo-Yu Xian$^{b,c}$\footnote{\email{zhuo-yu.xian@fu-berlin.de}}~,
Ren\'e Meyer$^{d,e}$\footnote{corresponding~author: {\email{rene.meyer@uni-wuerzburg.de}}}~,
Song He$^{f}$\footnote{corresponding~author: \email{hesong@nbu.edu.cn}}~~and
Bin Chen$^{f,g}$\footnote{corresponding~author: \email{chenbin1@nbu.edu.cn}}
}

\date{}
\maketitle
\vspace{-10mm}

\begin{center}
{
$^{a}$Center for Joint Quantum Studies and Department of Physics, School of Science, Tianjin University,\\
135 Yaguan Road, Tianjin 300350, China
\vspace{1.5mm}

$^{b}$School of Physics and Astronomy, Sun Yat-Sen University, 2 Daxue Road, Zhuhai 519082, China
\vspace{1.5mm}

$^{c}$Department of Physics, Freie Universit\"at Berlin, Arnimallee 14, DE-14195 Berlin, Germany
\vspace{1.5mm}

$^{d}$Institute for Theoretical Physics and Astrophysics and W\"urzburg-Dresden Cluster of Excellence ctd.qmat,
Julius-Maximilians-Universit\"at W\"urzburg, 97074 W\"urzburg, Germany
\vspace{1.5mm}

$^{e}$Shanghai Institute for Mathematics and Interdisciplinary Sciences (SIMIS),\\657 Songhu Road, Shanghai 200433, China
\vspace{1.5mm}

$^{f}$Institute of Fundamental Physics and Quantum Technology \& School of Physical Science and Technology,
Ningbo University, 818 Fenghua Road, Ningbo, Zhejiang 315211, China
\vspace{1.5mm}

$^{g}$School of Physics and Center for High Energy Physics, Peking University,\\
5 Yiheyuan Road, Beijing 100871, China
\vspace{1.5mm}
}
\vspace{10mm}
\end{center}

\begin{abstract}

  We propose a new product entropy, defined as the R\'enyi (or von Neumann) entropy of a normalized product operator constructed from two density matrices. We establish a duality showing that the SVD entanglement entropy of a subsystem for two pure states is exactly equivalent to the product entropy of the complementary subsystem. This connection provides both a transparent physical interpretation in terms of the spectral diversity of the subsystem state product and a computationally efficient route that bypasses the reduced transition matrix. For low-lying eigenstates, we derive analytical expressions for the subsystem product entropy between the ground state and primary excitations in two-dimensional conformal field theories, explicitly verified against the critical Ising chain. In quantum quench dynamics, the quasiparticle picture yields time evolution in the scaling limit: following a global quench, the subsystem product entropy exhibits distinct sequences of thermalization and revivals, whereas under a local operator quench, it develops characteristic plateaus whose constant values are determined by the inserted operator. Extensive numerical calculations on the critical Ising chain confirm the analytical predictions with excellent accuracy.

\end{abstract}

\baselineskip 18pt
\thispagestyle{empty}
\newpage

%\begin{center}
%\textbf{\LARGE Subsystem overlap entropy}\\~\\
%\today
%\end{center}

\tableofcontents

\section{Introduction}

Entanglement entropy has become a pivotal tool in the study of many-body quantum systems \cite{Amico:2007ag,Calabrese:2009bph,Laflorencie:2015eck}, quantum field theories \cite{Casini:2009sr,Witten:2018lha}, and quantum gravity \cite{Nishioka:2009un,Rangamani:2016dms}. For a system in pure state \(|\psi\rangle\) and a bipartition into a subsystem \(A\) and its complement \(B\), the reduced density matrix (RDM) \(\r_A = \tr_B |\psi\rangle\langle\psi|\) encodes the entanglement between the two parts of the total system. The most commonly used entanglement measures are the R\'enyi entropies
\be \label{SnrA}
S^{(n)}(\r_A) = - \frac{1}{n-1}\log\tr(\r_A^n),
\ee
and the von Neumann entropy
\be \label{S1rA}
S^{(1)}(\r_A) = -\tr(\r_A \log \r_A).
\ee
These quantities have been extensively investigated in ground states of critical systems, where conformal field theory (CFT) techniques provide universal predictions for ground states \cite{Calabrese:2004eu}, states out of equilibrium after quantum quenches~\cite{Calabrese:2005in}, and in low-lying excited states \cite{Alcaraz:2011tn,Berganza:2011mh}.

Although the entanglement of a single state is well understood, it is also important to compare two different quantum states and quantify their distinguishability or overlap. Traditional information-theoretic measures such as the fidelity \cite{Nielsen:2010oan,Watrous:2018rgz}
\be \label{Frs}
F(\r,\s) = \tr\sqrt{\sqrt{\r}\s\sqrt{\r}},
\ee
and the related Bures distance
\be
D(\r,\s) = \sqrt{2[1-F(\r,\s)]},
\ee
have proven useful in detecting quantum phase transitions~\cite{Parez:2022sgc} and in holographic settings~\cite{Kirklin:2019ror,Xiao:2023hnh,Sui:2025qve}. More recently, a new measure, termed SVD (singular value decomposition) entanglement entropy, has been introduced~\cite{Parzygnat:2023avh} which is constructed from the singular values of the reduced transition matrix \(\r_{A,\phi,\psi} = \tr_B |\phi\rangle\langle\psi|\) between two different pure states \(|\phi\rangle\) and \(|\psi\rangle\). For a particular choice of the exponent parameter, this SVD R\'enyi entropy reduces to the Alter-Brown-Botstein (ABB) entropy introduced earlier in the context of genome-wide expression data~\cite{Alter:2000wbf} and further analyzed and applied to random matrix theory as well as non-hermitian systems~\cite{Chen:2025ibe}. These quantities capture a notion of correlation between the two states that goes beyond what is contained in their individual RDMs.

In this paper, we introduce and study the {\it subsystem product entropy}, a quantity that generalizes the above constructions. We define a new product entropy \(S^{(n,m)}(\r,\s)\) for two arbitrary (pure or mixed) states \(\r\) and \(\s\) by first forming a normalized product operator \(P^{(m)}(\r,\s) \propto (\sqrt{\r}\,\s\sqrt{\r})^{m/2}\) and then computing the R\'enyi (or von Neumann) entropy of this operator. This definition is closely connected to the generalized fidelity \(F^{(m)}(\r,\s) = \tr[(\sqrt{\r}\s\sqrt{\r})^{m/2}]\), which interpolates between the standard fidelity at \(m=1\) and other measures for general \(m\)~\cite{Kirklin:2019ror,Parez:2022sgc,Xiao:2023hnh,Sui:2025qve}.
The product entropy admits a clear physical interpretation, and it quantifies the diversity of the overlap spectrum between the two states. It characterizes the diversity of the overlap component and does not depend on how large or small the overlap is, provided the overlap is nonvanishing.
In the special case \(\r = \s\), the product entropy reduces to a {\it rescaled R\'enyi entropy} of a single state, with an additional parameter \(m\) controlling the weight given to small or large eigenvalues. A key observation, which establishes the link to previous work, is that the SVD entanglement entropy of a subsystem \(A\) for two pure states \cite{Parzygnat:2023avh} is precisely equal to the product entropy of the complementary subsystem \(B\). This duality provides not only a transparent conceptual framework but also a computationally efficient route. Instead of working with the non-hermitian reduced transition matrix of subsystem $A$ for the two pure states, one can simply compute the product entropy of the two RDMs of subsystem $B$ corresponding to these same two pure states.

We apply the subsystem product entropy to a series of physically relevant settings. First, we examine low-lying eigenstates in a two-dimensional CFT and in the critical spin chain. Using the replica trick and conformal mappings, we derive explicit analytical expressions for the subsystem product entropy between the ground state and primary excited states. These CFT results are compared with exact numerical results for the critical Ising chain with periodic boundary conditions, showing perfect agreement up to finite-size corrections.

We then turn to the nonequilibrium dynamics of the product entropy after quantum quenches. Two distinct protocols are considered: a global quench \cite{Calabrese:2005in,Calabrese:2006rx,Calabrese:2016xau}, where the Hamiltonian is suddenly changed at time zero, and a local operator quench \cite{Nozaki:2014hna,Nozaki:2014uaa,He:2014mwa}, where the system is locally excited by the insertion of an operator in the initial state.
For the global quench on the Ising chain, we prepare initial states in both the Neveu--Schwarz (NS) and Ramond (R) sectors of the pre-quench Hamiltonian and evolve them with the critical Hamiltonian. We compute the time evolution of the subsystem Bures distance and the subsystem product entropy for a finite interval, using a linearized dispersion relation to highlight the quasiparticle-propagation picture. This linearization effectively removes irrelevant operators arising from UV band curvature, which is useful for our numerical simulations to converge to CFT predictions more efficiently \cite{Zhang:2019kwu}, c.f. additional discussions on this point in section 5.
For the local operator quench, we employ a powerful quasiparticle approach developed in CFT \cite{Nozaki:2014hna,Nozaki:2014uaa,He:2014mwa,Zhang:2019kwu}. When a local operator is inserted in the ground state, a pair of counter-propagating chiral quasiparticles is created, and their trajectories determine how the RDMs of a subsystem change over time. We classify the possible time regimes and provide analytic predictions for the subsystem Bures distance and product entropy.
For the specific inserted operators we consider, the subsystem product entropy exhibits plateau structures whose values are determined by the chiral weights the inserted operator. We test these predictions against numerical results on the critical Ising chain, using both a non-chiral operator insertion and a chiral one, again finding excellent agreement with the quasiparticle picture up to finite-size corrections.

The paper is organized as follows.
In Section~\ref{sectionOE} we define the product entropy, discuss its relation to the SVD entanglement entropy and the generalized fidelity, and analyze its properties.
Section~\ref{sectionLLE} presents the analytical CFT results of the subsystem product entropy for low-lying eigenstates and their numerical verification on the Ising chain.
Section~\ref{sectionQQ} contains our results on quench dynamics: after a brief review of the setup, we discuss global quenches in both NS and R sectors and local operator quenches for both non-chiral and chiral insertions.
We conclude with a summary and outlook in Section~\ref{sectionCon}.

\section{product entropy} \label{sectionOE}

In this section, we first present the definition of the product entropy and establish a connection between the subsystem product entropy of two pure states and the SVD entanglement entropy introduced in \cite{Parzygnat:2023avh}. The SVD entropy can be further generalized to mixed transition matrices following  \cite{guo2023constructible,parzygnat2023time}. In addition, we elaborate on the physical interpretation, monotonicity property, and measurement protocol of the product entropy.

\subsection{Definition}

For a quantum state described by the density matrix $\rho$, the standard R\'enyi and von Neumann entropies are defined in \eqref{SnrA} and \eqref{S1rA}, respectively. Consider two non-orthogonal density matrices $\rho$ and $\sigma$, satisfying $\operatorname{tr}(\rho\sigma)\neq 0$. We first introduce the normalized product
\be \label{Pmrs}
P^{(m)}(\rho,\sigma) = \frac{(\sqrt{\rho}\,\sigma\sqrt{\rho})^{m/2}}{\operatorname{tr}[(\sqrt{\rho}\,\sigma\sqrt{\rho})^{m/2}]},
\ee
and use it to define the product entropy via
\bea \label{Snmrs}
&& S^{(n,m)}(\rho,\sigma) = - \frac{1}{n-1}\log \operatorname{tr}\{ [ P^{(m)}(\rho,\sigma) ]^n \}, \nn\\
&& S^{(1,m)}(\rho,\sigma) = - \operatorname{tr}[ P^{(m)}(\rho,\sigma) \log P^{(m)}(\rho,\sigma) ].
\eea
We refer to the first expression as the R\'enyi product entropy and the second as the von Neumann product entropy.
As an alternative definition, one may take the normalized product to be
\be
P^{(m)}(\rho,\sigma) = \frac{(\rho\sigma)^{m/2}}{\operatorname{tr}[(\rho\sigma)^{m/2}]}.
\ee
This alternative formulation yields identical results for the R\'enyi and von Neumann product entropies.

A noteworthy special case occurs when the two density matrices coincide exactly. In this case, we have
\be
P^{(m)}(\rho,\rho) = \frac{\rho^m}{\operatorname{tr}(\rho^m)},
\ee
which may be interpreted as a rescaled version of $\rho$. Substituting this into the general definition yields the rescaled R\'enyi entropy and rescaled von Neumann entropy
\bea
&& S^{(n,m)}(\rho,\rho) = - \frac{1}{n-1}\log \frac{\operatorname{tr}(\rho^{nm})}{[\operatorname{tr}(\rho^m)]^n}, \nn\\
&& S^{(1,m)}(\rho,\rho) = - \operatorname{tr}\Big[ \frac{\rho^m}{\operatorname{tr}(\rho^m)} \log \frac{\rho^m}{\operatorname{tr}(\rho^m)} \Big]. \label{S1mrr}
\eea
The parameter $m\in[0,+\infty)$ acts as a weighting factor for the eigenvalues of $\rho$. Small values of $m$ assign greater weight to the smaller eigenvalues of $\rho$, while large values of $m$ emphasize the larger eigenvalues.

Consider a bipartite quantum system partitioned into a subsystem $A$ and its complement $B$. For two non-orthogonal pure states $|\phi\rag$ and $|\psi\rag$, one may define the reduced transition matrix as \cite{Nakata:2020luh}
\be
\r_{A,\phi,\psi} = \tr_B |\phi\rag \lag\psi|.
\ee
One can normalize this operator to obtain
\be
\t^{(m)}_{A,\phi,\psi} = \f{(\r_{A,\phi,\psi}^\dag\r_{A,\phi,\psi})^{m/2}}{\tr_A[(\r_{A,\phi,\psi}^\dag\r_{A,\phi,\psi})^{m/2}]}.
\ee
Note that the adjoint of the reduced transition matrix satisfies $\r_{A,\phi,\psi}^\dag=\r_{A,\psi,\phi}$. From this normalized operator, one arrives at the SVD entropy \cite{Parzygnat:2023avh}, given by
\bea \label{SVDE}
&& S^{(n,m)}_{A,\phi,\psi} = - \f{1}{n-1}\log \tr_A [ ( \t^{(m)}_{A,\phi,\psi} )^n ], \nn\\
&& S^{(1,m)}_{A,\phi,\psi} = - \tr_A [ \t^{(m)}_{A,\phi,\psi} \log \t^{(m)}_{A,\phi,\psi} ].
\eea
The special $m=1$ case was intensively studied in \cite{Parzygnat:2023avh}. For $m=2$, this reduces to the ABB entropy \cite{Alter:2000wbf,Chen:2025ibe}.

By comparing Eqs.~\eqref{Snmrs} and \eqref{SVDE}, we demonstrate that the SVD entropy of subsystem $A$ is precisely identical to the subsystem product entropy of its complementary subsystem $B$, as formalized in the relations
\bea
S^{(n,m)}_{A,\phi,\psi} &= S^{(n,m)}(\rho_{B,\phi},\rho_{B,\psi}), \nn\\
S^{(1,m)}_{A,\phi,\psi} &= S^{(1,m)}(\rho_{B,\phi},\rho_{B,\psi}).
\eea
To establish the desired equivalence, it is sufficient to demonstrate that the identity
\be
\tr_A \left[ \left( \r_{A,\psi,\phi} \r_{A,\phi,\psi} \right)^{p} \right] = \tr_B \left[ \left( \r_{B,\phi} \r_{B,\psi} \right)^{p} \right]
\ee
holds for all positive integers $p$, and extends via analytic continuation to all positive real $p$.
For a positive integer $p$, this relation admits a natural diagrammatic representation
\begin{center}
\includegraphics[width=0.7\textwidth]{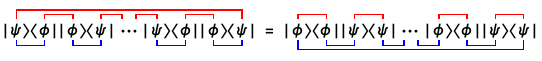}
\end{center}
\vspace{-4mm}
The upper red index contractions pertain to subsystem $A$, while the blue lower ones pertain to subsystem $B$.
This equivalence yields a computational advantage: to compute the SVD entropy for subsystem $A$, it is computationally more convenient and efficient to instead evaluate the subsystem product entropy of its complement $B$. {When either or both of the total system states $\rho$ and $\sigma$ are mixed, for instance, thermal states, the subsystem product entropy $S^{(n,m)}(\rho_B,\sigma_B)$ can no longer be interpreted as the SVD entanglement entropy of the complementary subsystem $A$. For generic mixed states $\r$ and $\s$ on the total system, the SVD entanglement entropy for subsystem $A$ is determined by the spectrum of the operator
\be
\f{[\tr_B(\s\r)\tr_B(\r\s)]^{m/2}}{\tr_A \big\{ [\tr_B(\s\r)\tr_B(\r\s)]^{m/2} \big\} },
\ee
whereas the subsystem product entropy for $B$ is determined by the spectrum of
\be
\f{(\r_B\s_B)^{m/2}}{\tr_B\left[(\r_B\s_B)^{m/2}\right]}.
\ee
In general, these two operators do not have identical spectra.}

For two quantum states with density matrices $\r$ and $\s$, one may define the generalized fidelity as \cite{Kirklin:2019ror,Parez:2022sgc,Xiao:2023hnh,Sui:2025qve}
\be\label{Fidelity}
F^{(m)}(\r,\s) = \tr[(\sqrt{\r}\s\sqrt{\r})^{m/2}] = \tr[(\r\s)^{m/2}].
\ee
For $m=1$, this quantity reduces to the standard quantum fidelity $F(\r,\s) = F^{(1)}(\r,\s)$, as presented in Eq.~\eqref{Frs}. The product entropy introduced in Eq.~\eqref{Snmrs} is directly linked to this generalized fidelity, via the explicit relations
\bea\label{SnmrsF}
&& S^{(n,m)}(\r,\s) = - \f{1}{n-1} \log \f{F^{(nm)}(\r,\s)}{\left[F^{(m)}(\r,\s)\right]^n},  \nn\\
&& S^{(1,m)}(\r,\s) = \log F^{(m)}(\r,\s) - \f{m}{F^{(m)}(\r,\s)} \f{\dd F^{(m)}(\r,\s)}{\dd m} .
\eea

We interpret the product entropy $S^{(1,m)}(\r,\s)$, defined in Eq.~\eqref{Snmrs}, as a quantifier of the spectral diversity of the overlap operator \eqref{Pmrs} between the quantum states $\r$ and $\s$, as varying $m$ emphasizes the influence of different parts of the spectrum of \eqref{Pmrs}. This interpretation remains valid for all pairs of states with non-vanishing overlap, regardless of the magnitude of the overlap.

When $\r$ and $\s$ are orthogonal, which means $\tr(\r\s)=0$, the normalized product given in Eq.~\eqref{Pmrs}, and by extension the product entropy in Eq.~\eqref{Snmrs}, are mathematically ill-defined. Moreover, for nearly orthogonal states $\r$ and $\s$, the product entropy exhibits severe numerical instability, making it an ill-defined quantity in this near-orthogonal regime as well.

\subsection{Monotonicity property}

Let the positive semi-definite matrix $\sqrt{\sqrt{\r}\s\sqrt{\r}}$ have non-vanishing eigenvalues $\{\a_i\}$. The product entropy can then be written in its spectral representation as
\be
S^{(1,m)}(\r,\s) = - \sum_i \Big( \f{\a_i^m}{\sum_j \a_j^m} \log \f{\a_i^m}{\sum_k \a_k^m} \Big).
\ee
Taking the derivative of this expression with respect to $m$ yields the strictly negative result
\be
\f{\dd }{\dd m} S^{(1,m)}(\r,\s) = - \f{m}{2(\sum_k \a_k^m)^2} \sum_{i,j} \Big[ \a_i^m \a_j^m \Big( \log \f{\a_i}{\a_j} \Big)^2 \Big] < 0.
\ee
This strict negativity establishes that the product entropy $S^{(1,m)}(\r,\s)$ is a monotonically decreasing function of the parameter $m$ over the interval $m\in[0,+\infty)$.

For $m=0$, the product entropy reduces to
\be
S^{(1,0)}(\r,\s) = \log a,
\ee
where $a$ denotes the number of non-vanishing eigenvalues of $\sqrt{\r}\s\sqrt{\r}$. In the limit $m\to+\infty$, we obtain
\be
S^{(1,+\infty)}(\r,\s) = \log b,
\ee
with $b$ being the degeneracy of the largest eigenvalue of $\sqrt{\r}\s\sqrt{\r}$. Analogous to the role of $m$ in the rescaled von Neumann entropy defined in Eq.~\eqref{S1mrr}, the parameter $m\in[0,+\infty)$ acts as a weight for the eigenvalues of $\sqrt{\r}\s\sqrt{\r}$. Small values of $m$ assign a greater weight to the smaller eigenvalues, while large values of $m$ prioritize its larger eigenvalues.

\subsection{Measurement} \label{measurement}

We now discuss how to measure the subsystem overlap entropy. The R\'enyi product entropy\eqref{SnmrsF} is determined by the two generalized fidelities in \eqref{Fidelity}. Since the operator $\sqrt{\rho_A}\sigma_A\sqrt{\rho_A}$ is positive semidefinite, $F^{(m)}(\rho_A,\sigma_A)$ is real and nonnegative for every $m>0$.

The most direct finite-copy measurement protocol applies when $m$ is even. Let $m=2r$, with $r\in\mathbb Z_{>0}$. Since $\rho_A\sigma_A$ and $\sqrt{\rho_A}\sigma_A\sqrt{\rho_A}$ have the same nonzero eigenvalues, one can equivalently write the even-index generalized fidelities in the R\'enyi product entropy\eqref{SnmrsF} as
\be\label{Fidelity_q}
F^{(2q)}(\rho_A,\sigma_A)
=
\tr_A[(\rho_A\sigma_A)^q],
\ee
with $q=r$ or $q=nr$. Thus, for integer R\'enyi index $n=2,3,\cdots$, the product entropyat $m=2r$ is obtained from the two moments
\be\label{SnmrsFeven}
S^{(n,2r)}(\rho_A,\sigma_A)
=
-\frac{1}{n-1}
\log
\frac{F^{(2nr)}(\rho_A,\sigma_A)}
{\big[ F^{(2r)}(\rho_A,\sigma_A) \big]^n}.
\ee

Let $V_A^{(k)}$ denote the cyclic shift acting only on the $k$ copies of subsystem $A$ and trivially on the complements $\bar A$, namely,
\be
\tr\!\left[
V_A^{(k)}
\left(
X_1\otimes X_2\otimes\cdots\otimes X_k
\right)
\right]
=
\tr_A(X_{1,A}X_{2,A}\cdots X_{k,A}),
\ee
where we have used $X_{j,A}=\tr_{\bar A}X_j$ for $j=1,2,\cdots,k$.
Then the even-index generalized fidelity is written as
\be \label{cyclemeasurement}
F^{(2q)}(\rho_A,\sigma_A)
=
\tr\!\big[
V_A^{(2q)}
(
\rho\otimes\sigma
)^{\otimes q}
\big].
\ee
Here the $2q$ copies are prepared in the alternating order $\rho,\sigma,\rho,\sigma,\cdots,\rho,\sigma,$
and only the $A$-parts of the replicas are cyclically permuted. The complementary degrees of freedom are left untouched and traced out. This is the same permutation logic that underlies direct measurements of nonlinear functionals and R\'enyi entropies \cite{Ekert:2002qtj,Daley:2012xhf,Islam:2015mom}.

\begin{figure}[t]
\centering
%\begin{tikzpicture}[x=1cm,y=0.78cm,>=Latex,font=\small]
\begin{tikzpicture}[x=1cm,y=0.78cm,font=\small]
\tikzset{
wire/.style={line width=0.55pt},
gate/.style={draw,minimum width=0.9cm,minimum height=0.36cm,inner sep=1pt,fill=white},
prep/.style={draw,minimum width=0.7cm,minimum height=0.82cm,inner sep=1pt,align=center,fill=white},
meterbox/.style={draw,minimum width=0.6cm,minimum height=0.42cm,inner sep=0pt,fill=white}
}

\draw[wire] (0,0) -- (6,0);
\foreach \y in {-1,-1.45,-2.2,-2.65,-3.4,-3.85,-4.6,-5.05}
  \draw[wire] (0,\y) -- (6,\y);

\node[anchor=east] at (-0.10,0) {$|0\rangle_a$};
\node[anchor=east] at (-0.10,-1) {$A_1$};
\node[anchor=east] at (-0.10,-1.45) {$\bar A_1$};
\node[anchor=east] at (-0.10,-2.2) {$A_2$};
\node[anchor=east] at (-0.10,-2.65) {$\bar A_2$};
\node[anchor=east] at (-0.10,-3.4) {$A_3$};
\node[anchor=east] at (-0.10,-3.85) {$\bar A_3$};
\node[anchor=east] at (-0.10,-4.6) {$A_4$};
\node[anchor=east] at (-0.10,-5.05) {$\bar A_4$};

\node[prep] at (1,-1.225) {$\rho$};
\node[prep] at (1,-2.425) {$\sigma$};
\node[prep] at (1,-3.625) {$\rho$};
\node[prep] at (1,-4.825) {$\sigma$};

\node[gate] at (1,0) {$H$};

\filldraw (2.5,0) circle (1.8pt);
\draw[wire] (2.5,0) -- (2.5,-4.6);
\foreach \y in {-1,-2.2,-3.4,-4.6}
  \node[gate] at (2.5,\y) {$V_A^{(4)}$};

\node[anchor=west,align=left] at (1.5,-5.5)
  {cyclic shift on $A_1A_2A_3A_4$};

\node[gate] at (4,0) {$H$};

\node[meterbox] (ma) at (5.7,0) {};
\draw[line width=0.55pt]
  ($(ma.center)+(-0.2,-0.05)$)
  arc[start angle=170,end angle=10,radius=0.2cm];
%\draw[-Latex,line width=0.55pt]
\draw[line width=0.55pt]
  ($(ma.center)+(0,-0.05)$) -- ($(ma.center)+(0.2,0.15)$);

\node[anchor=west] at (6,0) {$Z_a$};

% \draw[decorate,decoration={brace,mirror,amplitude=4pt},thick]
%   (-1,-0.82) -- (-1,-5.23)
%   node[midway,left=6pt,align=center] {four copies\\$\rho,\sigma,\rho,\sigma$};

\end{tikzpicture}
\caption{Hadamard-test measurement of the numerator moment $F^{(4)}(\rho_A,\sigma_A)=\tr_A[(\rho_A\sigma_A)^2]$. Four alternating copies
$\rho,\sigma,\rho,\sigma$ are prepared, and the controlled cyclic shift $V_A^{(4)}$ acts only on the subsystem replicas $A_1A_2A_3A_4$. The final computational-basis measurement gives $\langle Z_a\rangle=\mathrm{Re}\,F^{(4)}(\rho_A,\sigma_A)$, which equals $F^{(4)}(\rho_A,\sigma_A)$ in the ideal case. The denominator uses similar circuit with two copies, $\rho,\sigma$, and $V_A^{(4)}$ replaced by the subsystem swap $V_A^{(2)}$, giving $F^{(2)}(\rho_A,\sigma_A)=\tr_A(\rho_A\sigma_A)$. Hence, we obtain $S^{(2,2)}(\r_A,\s_A)=-\log\{F^{(4)}(\rho_A,\sigma_A)/[F^{(2)}(\rho_A,\sigma_A)]^2\}$.}
\label{fig:measurement-m2n2}
\end{figure}
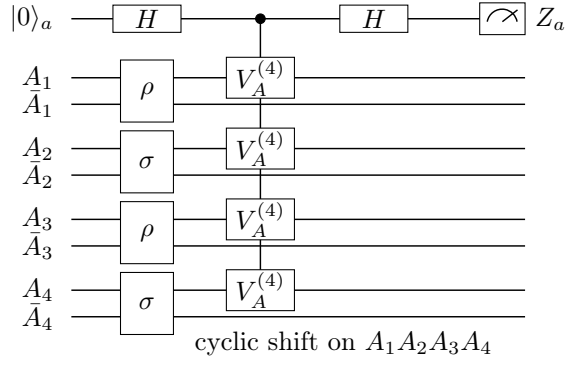

The even-index generalized fidelity $F^{(2q)}(\rho_A,\sigma_A)$ can be obtained from a Hadamard test with a controlled cyclic shift (cf. Fig.~\ref{fig:measurement-m2n2}). The circuit prepares the ancilla in $|0\rangle_a$, applies a Hadamard gate $H=\tfrac{1}{\sqrt2}\bigl(\begin{smallmatrix}1&1\\1&-1\end{smallmatrix}\bigr)$, and then applies the controlled unitary $U=|0\rangle\langle0|_a\otimes \mathbf{1}+|1\rangle\langle1|_a\otimes V_A^{(2q)}$. Before the final ancilla measurement, the RDM of the ancilla is
\be
\rho_a
=
\frac12
\begin{pmatrix}
1 & \tr\!\big[(\rho\otimes\sigma)^{\otimes q}(V_A^{(2q)})^\dagger\big] \\
\tr\!\big[V_A^{(2q)}(\rho\otimes\sigma)^{\otimes q}\big] & 1
\end{pmatrix}.
\ee
Consequently, with \eqref{cyclemeasurement}, for the Pauli operators $X_a$, $Y_a$, and $Z_a$ of the ancilla, one obtains
\be
\langle X_a\rangle_{\rho_a}
=
\mathrm{Re}\,F^{(2q)}(\rho_A,\sigma_A)
\overset{\text{ideal}}{=}
F^{(2q)}(\rho_A,\sigma_A),
~~
\langle Y_a\rangle_{\rho_a}
=
\mathrm{Im}\,F^{(2q)}(\rho_A,\sigma_A)
\overset{\text{ideal}}{=}
0.
\ee
where $\overset{\text{ideal}}{=}$ indicates equality in the ideal, noiseless limit, where the Hadamard test is implemented without errors. A nonzero imaginary part of the generalized fidelity in measurements therefore provides a useful diagnostic of non-ideal implementation, such as coherent gate errors, imperfect state preparation or measurement errors, and finite-shot statistical fluctuations. For more general ordered products of several noncommuting density matrices, $\tr_A(\rho_{1,A}\rho_{2,A}\cdots\rho_{k,A})$, the trace need not be real, and both the real and imaginary quadratures are then physically required.

To measure the real part, one applies a final Hadamard gate and then measures the ancilla in the computational basis. The measured expectation value is
\be
\langle Z_a\rangle_{H\rho_aH^\dagger}
=
\langle X_a\rangle_{\rho_a}
=
\mathrm{Re}\,F^{(2q)}(\rho_A,\sigma_A).
\ee
For the present alternating overlap moments, this quantity should be nonnegative up to statistical and experimental errors. To measure the imaginary part, one inserts gate
$
S^\dagger =\bigl(
\begin{smallmatrix}
1 & 0 \\
0 & -\ii
\end{smallmatrix}\bigr)
$
before the final Hadamard, and also measures the ancilla in the computational basis. The measured expectation value is
\be
\langle Z_a\rangle_{HS^\dagger\rho_a S H^\dagger}
=
\langle Y_a\rangle_{\rho_a}
=
\mathrm{Im}\,F^{(2q)}(\rho_A,\sigma_A).
\ee
For the present alternating overlap moments, this quantity should vanish up to statistical and experimental errors.
Thus, the Hadamard test provides a direct measurement protocol for $F^{(2q)}(\rho_A,\sigma_A)$, from which the product entropy
$S^{(n,2r)}(\rho_A,\sigma_A)$ follows by the ratio in \eqref{SnmrsFeven}. The quantum circuit in Fig.~\ref{fig:measurement-m2n2} displays the simplest nontrivial case for $F^{(4)}(\rho_A,\sigma_A)$.

Several experimental variants can implement the even-$m$ estimator. If coherent multi-copy control is available, one can use controlled cyclic shifts as above. Alternatively, destructive many-body interference between replicated systems can access permutation observables, as in cold-atom measurements of purity and second R\'enyi entropy \cite{Daley:2012xhf,Islam:2015mom}. If coherent permutations are not available, randomized-measurement protocols provide a complementary route: statistical correlations of local random bases can estimate purities, products, and related R\'enyi quantities using repeated preparations and classical post-processing \cite{Elben:2018xmb,Brydges:2019wut,Elben:2019dxb,Elben:2019yre}. For the present product entropy, such methods provide direct access to the lowest even generalized fidelity $F^{(2)}(\rho_A,\sigma_A)$, whereas higher moments require higher-order correlations or the reconstruction of the moment sequence.

Odd integer values of $m$ are qualitatively different. If $m=2r+1$, then
\be
F^{(2r+1)}(\rho_A,\sigma_A)
=
\tr_A\!\big[
(\sqrt{\rho_A}\sigma_A\sqrt{\rho_A})^{r+1/2}
\big].
\ee
This quantity is still real and nonnegative, but it contains a fractional power of the positive operator $\sqrt{\rho_A}\sigma_A\sqrt{\rho_A}$. It is therefore not an ordinary polynomial trace moment and cannot, in general, be reduced to a finite-copy cyclic permutation measurement. The same obstruction applies in the von Neumann product entropy $S^{(1,m)}(\r_A,\s_A)$ in \eqref{SnmrsF}. Accessing
these fidelities require reconstructing sufficient spectral information about $\sqrt{\rho_A}\sigma_A\sqrt{\rho_A}$, for example, through tomography, randomized-measurement reconstruction, classical shadows, or quantum spectral
or singular-value estimation \cite{Nielsen:2010oan,Elben:2018xmb,Elben:2019dxb,Huang:2020tih,Gilyen:2018khw}.

\section{Low-lying eigenstates} \label{sectionLLE}

In this section, we derive analytical results of subsystem product entropy between low-lying eigenstates in 2D CFT, specifically the 2D free massless fermion theory, and perform complementary numerical calculations for the critical Ising spin chain. We find excellent agreement between analytical predictions and numerical results.

\subsection{General 2D CFT}

We consider a 2D CFT defined on a circle of circumference $L$. For a spatial subsystem $A=[0,\ell]$ in the  ground state $|G\rag$, the R\'enyi moments of the RDM satisfy the well-known result \cite{Calabrese:2004eu}
\be\label{eq:CalabreseResult}
\tr_A \r_{A,G}^n = c_n \Big( \f{L}{\pi \e} \sin \f{\pi\ell}{L} \Big)^{-2\D_n},
\ee
where $c_n$ is a non-universal constant factor, $\e$ denotes the ultraviolet (UV) cutoff, $\D_n = \f{c}{12} \left( n - \f{1}{n} \right)$ is the scaling dimension of the corresponding twist operators, and $c$ is the central charge of the CFT. Using this relation, we derive the rescaled R\'enyi entropy for the ground state
\bea
&& S^{(n,m)}(\r_{A,G},\r_{A,G}) = \f{c(n+1)}{6mn} \log \Big( \f{L}{\pi\e} \sin \f{\pi\ell}{L} \Big) - \f{1}{n-1} \log \f{c_{mn}}{c_m^n}, \nn\\
&& S^{(1,m)}(\r_{A,G},\r_{A,G}) = \f{c}{3m} \log \Big( \f{L}{\pi\e} \sin \f{\pi\ell}{L} \Big) - \f{m c'_m}{c_m} + \log c_m.
\eea
Here $c_{mn}$ is the non-universal constant factor in $\tr_A \r_{A,G}^{mn}$.

We consider a general normalized primary excited state $|\phi\rag$ with scaling dimension $\D_\phi$ and conformal spin $s_\phi$ in the 2D CFT. Via conformal mapping, the following relation holds for the ratio of R\'enyi moments of the RDM \cite{Alcaraz:2011tn, Berganza:2011mh}
\be \label{FphinellL}
F_\phi\Big(n,\f{\ell}{L}\Big) \equiv \f{\tr_A \r_{A,\phi}^n}{\tr_A \r_{A,G}^n}
                              = \ii^{2ns_\phi} \Big( \f{2}{n} \sin \f{\pi\ell}{L} \Big)^{2n\D_\phi} G_\phi\Big(n,\f{2\pi\ell}{n L}\Big),
\ee
where $G_\phi(n,\th)$ denotes the $2n$-point correlation function on a complex plane, defined as
\be \label{Gphinth}
G_\phi(n,\th) \equiv \Big\lag \prod_{j=0}^{n-1}
\Big[  (f_j f_j^\th )^{h_\phi} (\bar f_j \bar f_j^\th )^{\bar h_\phi} \phi(f_j^\th,\bar f_j^\th) \phi^\dag(f_j,\bar f_j) \Big]
\Big\rag_C,
\ee
with the operator insertion points given by $f_j\equiv\ep^{\f{2\pi\ii j}{n}}$ and $f_j^\th\equiv\ep^{ \ii \left( \f{2\pi j}{n} + \th \right)}$. Using the function $F_\phi(n,\ell/L)$, we derive the difference  between the excited state and the ground state in rescaled R\'enyi entropy
\be \label{DSnmrAphirAphi}
\D S^{(n,m)}( \r_{A,\phi}, \r_{A,\phi} ) \equiv S^{(n,m)}( \r_{A,\phi}, \r_{A,\phi} ) - S^{(n,m)}( \r_{A,G}, \r_{A,G} )
                                         = - \f{1}{n-1} \log \f{F_\phi(mn,\f{\ell}{L})}{[F_\phi(m,\f{\ell}{L})]^n }.
\ee

Similarly, taking $m$ to be an even integer, we obtain the ratio of traces over subsystem $A$
\be
\f{\tr_A[(\r_{A,G}\r_{A,\phi})^{m/2}]}{\tr_A\r_{A,G}^m} = \ii^{m s_\phi} \Big( \f{2}{m} \sin\f{\pi\ell}{L} \Big)^{m\D_\phi} G_\phi\Big( \f{m}{2},\f{2\pi\ell}{m L} \Big),
\ee
where $G_\phi(n,\th)$ is the same $2n$-point correlation function on the complex plane defined in Eq.~\eqref{Gphinth}. From the definition of $F_\phi(n,\ell/L)$ in Eq.~\eqref{FphinellL}, we invert the relation to express the function $G_\phi$ in terms of the function $F_\phi$
\be
G_\phi(n,\th) = \ii^{-2n s_\phi} \Big( \f{2}{n} \sin \f{n\th}{2} \Big)^{-2n\D_\phi} F_\phi\Big( n, \f{n\th}{2\pi} \Big).
\ee
Substituting this identity into the trace ratio above, we arrive at the simplified closed-form expression
\be
\f{\tr_A[(\r_{A,G}\r_{A,\phi})^{m/2}]}{\tr_A\r_{A,G}^m} = \left( \f{\sin\f{\pi\ell}{L}}{2\sin\f{\pi\ell}{2L}} \right)^{m\D_\phi} F_\phi\Big( \f{m}{2}, \f{\ell}{2L} \Big).
\ee
Note that while the conformal spin $s_\phi$ appears in intermediate steps of this calculation, the final product entropy must be independent of it, as it is a real and positive quantity.
Using this result, we define the difference between the subsystem product entropy and the ground state rescaled R\'enyi entropy
\be \label{DSnmrAGrAphi}
\D S^{(n,m)}( \r_{A,G}, \r_{A,\phi} ) \equiv S^{(n,m)}( \r_{A,G}, \r_{A,\phi} ) - S^{(n,m)}( \r_{A,G}, \r_{A,G} )
                                         = - \f{1}{n-1} \log \f{F_\phi(\f{mn}{2},\f{\ell}{2L})}{[F_\phi(\f{m}{2},\f{\ell}{2L})]^n }.
\ee

We introduce two auxiliary functions to encode the rescaled R\'enyi entropy and subsystem product entropy, respectively
\bea \label{fphinmellL}
&& f_\phi\Big(n,m,\f{\ell}{L}\Big) \equiv \D S^{(n,m)}( \r_{A,\phi}, \r_{A,\phi} ), \nn\\
&& g_\phi\Big(n,m,\f{\ell}{L}\Big) \equiv \D S^{(n,m)}( \r_{A,G}, \r_{A,\phi} ).
\eea
By direct comparison of Eq.~\eqref{DSnmrAphirAphi} and Eq.~\eqref{DSnmrAGrAphi}, we establish the exact mapping between these two functions
\be
g_\phi\Big(n,m,\f{\ell}{L}\Big) = f_\phi\Big(n,\f{m}{2},\f{\ell}{2L}\Big).
\ee
This relation demonstrates that the single function $f_\phi(n,m,\f{\ell}{L})$ fully specifies both the excess rescaled R\'enyi entropy $\D S^{(n,m)}( \r_{A,\phi}, \r_{A,\phi} )$ and the excess subsystem product entropy $\D S^{(n,m)}( \r_{A,G}, \r_{A,\phi} )$.

\subsection{2D free massless fermion theory}

In 2D, the free massless fermion theory is a  CFT with central charge $c=\frac{1}{2}$. We focus on the ground state $|G\rag$, as well as excited states $|\psi\rag$ and $|\ve\rag$ generated by primary operators: the operator $\psi$ has conformal weights $(\frac{1}{2},0)$, while the operator $\ve$ has conformal weights $(\frac{1}{2},\frac{1}{2})$.

Using the definition of $F_\phi(n,\ell/L)$ in Eq.~\eqref{FphinellL}, the following exact result for the $\ve$ primary has been derived in the literature \cite{Essler:2012pni,Calabrese:2014ntv}
\be
F_\ve\Big(n,\f{\ell}{L}\Big) = \Big( \f2n \sin\f{\pi\ell}{L} \Big)^{2n}
\f{\Gamma^2\left(\f{1+n+n\csc\f{\pi\ell}{L}}{2}\right)}
  {\Gamma^2\left(\f{1-n+n\csc\f{\pi\ell}{L}}{2}\right)}.
\ee
With  the auxiliary function $f_\phi(n,m,\ell/L)$, we defined earlier in Eq.~\eqref{fphinmellL}, we derive the explicit closed-form expressions
\bea
&& f_\ve \Big(n,m,\f{\ell}{L}\Big) = \f{2mn}{n-1} \log n - \f{1}{n-1} \log
                                    \bigg[
                     \f{\Gamma^2(\f{1+ m n + m n \csc\f{\pi\ell}{L}}{2}) \Gamma^{2n}(\f{1-m + m \csc\f{\pi\ell}{L}}{2})}
                       {\Gamma^2(\f{1-m n + m n \csc\f{\pi\ell}{L}}{2}) \Gamma^{2n}(\f{1+m + m \csc\f{\pi\ell}{L}}{2})}
                                    \bigg] \nn\\
&& f_\ve \Big(1,m,\f{\ell}{L}\Big) = - m \Big( 1 + \csc\f{\pi\ell}{L} \Big) \digamma \Big( \f{1 + m + m \csc\f{\pi\ell}{L}}{2} \Big)
                                     - m \Big( 1 - \csc\f{\pi\ell}{L} \Big) \digamma \Big( \f{1 - m + m \csc\f{\pi\ell}{L}}{2} \Big) \nn\\
&& \phantom{f_\ve \Big(1,m,\f{\ell}{L}\Big) =}
                                     +2m + 2 \log \f{\Gamma(\f{1+m + m \csc\f{\pi\ell}{L}}{2})}
                                              {\Gamma(\f{1-m + m \csc\f{\pi\ell}{L}}{2})},
\eea
where $\digamma(x)=\Gamma'(x)/\Gamma(x)$ denotes the digamma function.
As established in the preceding section, the function $f_\ve(n,m,\ell/L)$ fully specifies both the excess rescaled R\'enyi entropy $\Delta S^{(n,m)}( \r_{A,\ve}, \r_{A,\ve} )$ and the excess subsystem product entropy $\Delta S^{(n,m)}( \r_{A,G}, \r_{A,\ve} )$. For the fermionic primary operator $\psi$, we note the simple relation
\be
F_\psi\Big(n,\f{\ell}{L}\Big) = \sqrt{F_\ve\Big(n,\f{\ell}{L}\Big)},
\ee
from which we directly obtain the corresponding excess rescaled R\'enyi entropy $\Delta S^{(n,m)}( \r_{A,\psi}, \r_{A,\psi} )$ and excess subsystem product entropy $\Delta S^{(n,m)}( \r_{A,G}, \r_{A,\psi} )$, encoded in the auxiliary function
\be
f_\psi \Big(n,m,\f{\ell}{L}\Big) = \f12 f_\ve \Big(n,m,\f{\ell}{L}\Big).
\ee

Examples of the rescaled R\'enyi entropy and subsystem product entropy for the free massless fermion theory are presented in Figures~\ref{FigureRRE} and \ref{FigureOE}.

\subsection{Critical Ising chain}

We consider a one-dimensional spin-1/2 transverse-field Ising chain, described by the Hamiltonian
\be \label{Hh}
H(h) = - \f12 \sum_{j=1}^L \left( \s^x_j \s^x_{j+1} + h \s^z_j \right),
\ee
where $\s^\mu_j$ ($\mu=x,y,z$) denote the Pauli matrices acting on the $j$-th lattice site, and the system is subject to periodic boundary conditions $\s^x_{L+1}=\s^x_1$. The continuous limit of the model at its quantum critical point, corresponding to the transverse field $h=1$, maps exactly to the two-dimensional free massless fermion theory \cite{Mussardo:2010mbh}.

\begin{figure}[t]
  \centering
  \includegraphics[width=0.9\textwidth]{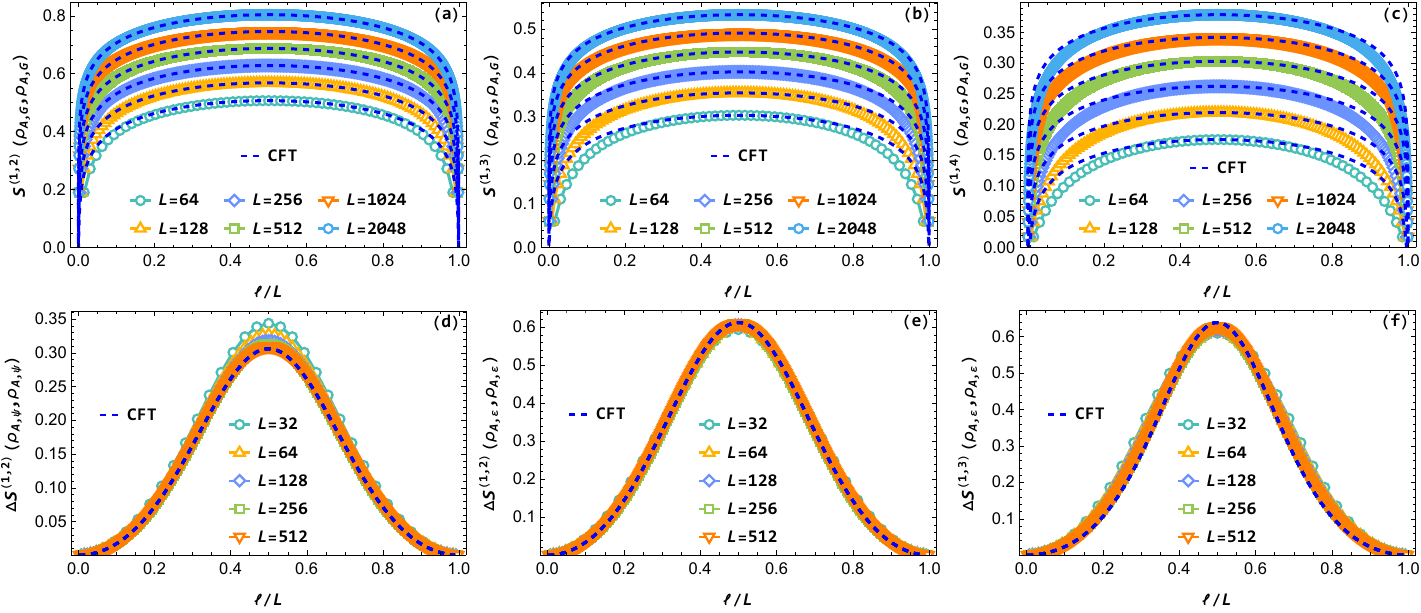}\\
  \caption{Rescaled R\'enyi entropy for low-lying eigenstates of the 2D free massless fermion theory (dashed blue lines) and the critical Ising chain (symbols). For the top row, a constant offset has been applied to the CFT results to align them with the spin chain data at $\ell=\f{L}{2}$. In general, there is a difference in the non-universal part given by $c_n$ in \eqref{eq:CalabreseResult} between numerical spin chain results and the CFT results. This difference is, however, the same for fixed $L$, and hence can be removed by a $L$-dependent offset.}
  \label{FigureRRE}
\end{figure}

\begin{figure}[t]
  \centering
  \includegraphics[width=0.9\textwidth]{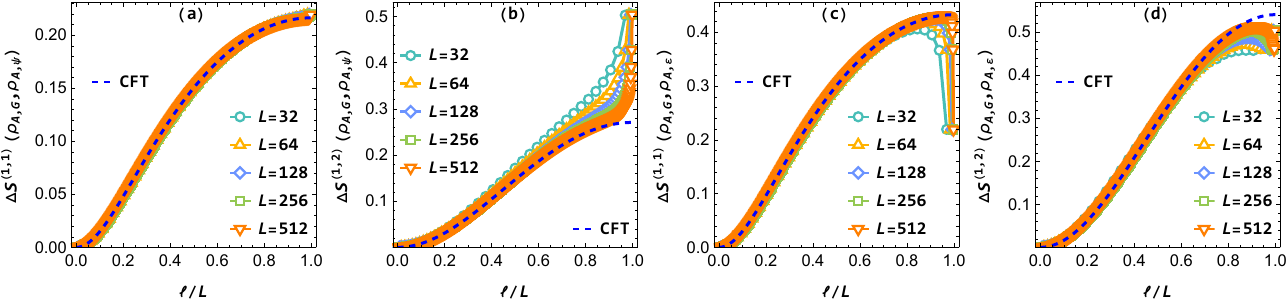}\\
  \caption{Subsystem product entropy between low-lying eigenstates, with analytical results from the 2D free massless fermion theory (dashed blue lines) and numerical results from the critical Ising chain (symbols). %\blue{RM: Why does the large $\ell/L$ data deviate from the CFT result so much in figures b and c, but less in d and nearly not in a?} \magenta{I have no idea why. I will check the numerical data to make sure no mistakes here.}
  }
  \label{FigureOE}
\end{figure}

We present numerical results for the rescaled R\'enyi function and subsystem product entropy of the critical Ising chain in Figures~\ref{FigureRRE} and \ref{FigureOE}. In the scaling limit, we find perfect agreement between the analytical predictions of the 2D CFT and the numerical results for the lattice spin chain. We note that significant finite-size corrections emerge in the regimes $\ell\to1$ and $\ell\to L-1$, where the subsystem size approaches the lattice spacing or the full system size minus one lattice site.

\section{Quantum quench} \label{sectionQQ}

In this section, we investigate the subsystem product entropy in the dynamical evolution following a quantum quench, including both global and local quench protocols.

We start from a bipartite system partitioned into subsystem $A$ and its complementary subsystem $B$, with the total system evolving in the time-dependent pure state $|\psi(t)\rangle$. The RDMs for the two subsystems are defined as
\bea
&& \r_A(t) = \tr_B |\psi(t)\rag \lag\psi(t)|, \nn\\
&& \r_B(t) = \tr_A |\psi(t)\rag \lag\psi(t)|.
\eea

For subsystem $A$, we evaluate three core dynamical quantities: the entanglement entropy
\be
S_A(t) = S^{(1)}(\r_A(t)),
\ee
the time-dependent subsystem Bures distance
\be
D_A(t_1,t_2) = D( \r_A(t_1), \r_A(t_2) ),
\ee
and the subsystem product entropy
\be
S_A^{(1,m)}(t_1,t_2) = S^{(1,m)}(\r_A(t_1), \r_A(t_2)).
\ee
The corresponding quantities for the complementary subsystem $B$ are defined in an identical manner
\be
S_B(t) = S^{(1)}(\r_B(t)),
\ee
\be
D_B(t_1,t_2) = D( \r_B(t_1), \r_B(t_2) ),
\ee
\be
S_B^{(1,m)}(t_1,t_2) = S^{(1,m)}(\r_B(t_1), \r_B(t_2)).
\ee

\subsection{Global quench}

We adopt the standard global quench protocol introduced in \cite{Calabrese:2005in}. We consider a periodic Ising chain governed by the Hamiltonian defined in Eq. (\ref{Hh}). The system is initialized in the state $|\psi(0)\rangle$, which is prepared as an eigenstate of the pre-quench Hamiltonian $H(h_0)$. At $t=0$, we perform the quench, and the subsequent unitary evolution of the state is governed by the post-quench Hamiltonian $H(h)$ with $h \neq h_0$. The time-dependent state is given by
\be
|\psi(t)\rag = \ep^{-\ii H(h) t} |\psi(0)\rag.
\ee

\subsubsection{NS sector}

\begin{figure}[t]
  \centering
  \includegraphics[width=0.9\textwidth]{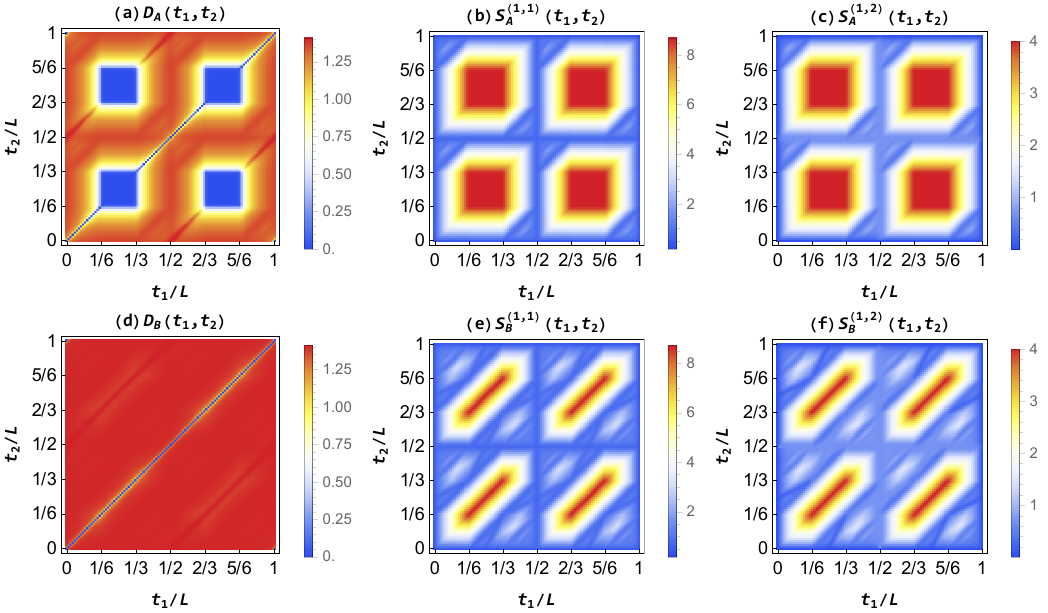}\\
  \caption{Time evolution of dynamical quantities following a global quench on a critical Ising chain of length $L$, initialized in a state from the NS sector. The first, second, and third columns display the subsystem Bures distance, subsystem product entropy with $m=1$, and subsystem product entropy with $m=2$, respectively. The first row presents results for subsystem $A=[1,\ell]$, and the second row presents results for its complementary subsystem $B=[\ell+1, L]$. All results are obtained with fixed parameters $h_0=2$, $h=1$, $L=120$, and $\ell=40$, using a linearized dispersion relation for the post-quench critical Ising chain. {Note that the post-quench Hamiltonian is critical, and our convention is such that the speed of quasiparticles post-quench is  $v=1$.}}
  \label{FigureGlobalQuench1}
\end{figure}

We first focus on the scenario where the initial state $|\psi(0)\rag$ is the ground state of the pre-quench Hamiltonian $H(h_0)$, corresponding to the empty state in the NS sector $|\NS\rag$. In Fig.~\ref{FigureGlobalQuench1}, we plot the time evolution of two core dynamical quantities: the subsystem Bures distance and the subsystem product entropy for $m=1$ and $m=2$. To better elucidate the underlying quasiparticle picture and benchmark our results against the analytical predictions of 2D CFT, we follow \cite{Zhang:2019kwu} and employ a linearized dispersion relation for the post-quench Hamiltonian $H(h)$ evaluated at $h=1$, given by
\be
\ve_k = 2 \sin\f{\pi |k|}{L} \to \f{2\pi |k|}{L}, ~ k \in \Big[ - \f{L}{2}, \f{L}{2} \Big).
\ee
As has been discussed in \cite{Zhang:2019kwu}, it is convenient to use the linearized dispersion relation so that the spin chain results more efficiently converge to the CFT predictions.

As shown in Fig.~\ref{FigureGlobalQuench1}, and in full agreement with the picture reported in Ref. \cite{Cardy:2014rqa}, the evolution of a subsystem $A$ of length $0<\ell<\f{L}{2}$ over a single period $0<t<L$ can be partitioned into six distinct stages {Following \cite{Cardy:2014rqa},  the panels a,b,c of Fig.~4 match their picture. The blue regions in panel a and the red regions in panel b and c correspond to the thermalized RDM.  The other time regimes correspond to the processes of thermalization or de-thermalization. Here, de-thermalization means that the subsystem approaches another non-thermal state, which is different from the original state the system started with.}
\bea
& 0 < t < \f{\ell}{2}               & \textrm{~process of thermalization}, \nn\\
& \f{\ell}{2} < t < \f{L-\ell}{2}   & \textrm{~staying thermalized}, \nn\\
& \f{L-\ell}{2} < t < \f{L}{2}      & \textrm{~process of de-thermalization (not revival)}, \nn\\
& \f{L}{2} < t < \f{L+\ell}{2}      & \textrm{~process of thermalization}, \nn\\
& \f{L+\ell}{2} < t < L-\f{\ell}{2} & \textrm{~staying thermalized}, \nn\\
& L-\f{\ell}{2} < t < L             & \textrm{~process of revival}.
\eea
Notably, the subsystem Bures distance and the subsystem product entropy exhibit opposite thermalization signatures.  Specifically, thermalization corresponds to a low value for the subsystem Bures distance, but a high value for the subsystem product entropy.

\subsubsection{R sector}

\begin{figure}[t]
  \centering
  \includegraphics[width=0.9\textwidth]{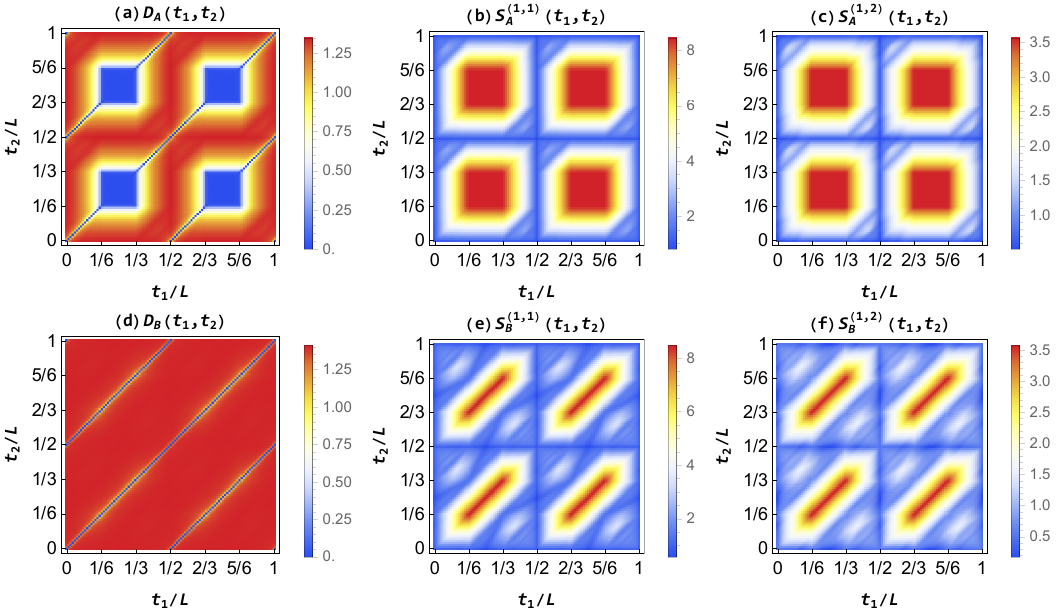}\\
  \caption{Time evolution of dynamical quantities following a global quench on a critical Ising chain of length $L$, initialized in a state from the R sector.  The first, second, and third columns display the subsystem Bures distance, subsystem product entropy with $m=1$, and subsystem product entropy with $m=2$, respectively. The first row presents results for subsystem $A=[1,\ell]$, and the second row presents results for its complementary subsystem $B=[\ell+1, L]$. All results are obtained with fixed parameters $h_0=2$, $h=1$, $L=120$, and $\ell=40$, using a linearized dispersion relation for the post-quench critical Ising chain.}
  \label{FigureGlobalQuench2}
\end{figure}

We next turn to the scenario where the initial state is $|\psi(0)\rag=c_0^\dag |\rR\rag$, which belongs to the R sector of the pre-quench Hamiltonian $H(h_0)$, where $|\rR\rag$ denotes the empty state in the R sector. In Fig.~\ref{FigureGlobalQuench2}, we plot the time evolution of the subsystem Bures distance, as well as the subsystem product entropy for $m=1$ and $m=2$. From the results, we can resolve the full dynamical evolution of a subsystem $A$ of size $0<\ell<\f{L}{2}$ over a single period $0<t<\f{L}{2}$, which can be divided into three distinct stages
\bea
& 0 < t < \f{\ell}{2}               & \textrm{~process of thermalization}, \nn\\
& \f{\ell}{2} < t < \f{L-\ell}{2}   & \textrm{~staying thermalized}, \nn\\
& \f{L-\ell}{2} < t < \f{L}{2}      & \textrm{~process of revival}.
\eea
Note the subtle period difference in the global quench dynamics between initial states prepared in the NS and R sectors, the reason for which is that the modes in the NS sector are half-integer while the R sector has integer modes.

\subsection{Local operator quench}

For the local operator quench setup, we follow the protocol introduced in Refs. \cite{Nozaki:2014hna,Nozaki:2014uaa,He:2014mwa}. Within this framework, we compute the subsystem Bures distance and the subsystem product entropy, with calculations carried out both in the 2D CFT and for the critical Ising spin chain.

\subsubsection{2D CFT}

\begin{figure}[t]
  \centering
  \includegraphics[width=0.7\textwidth]{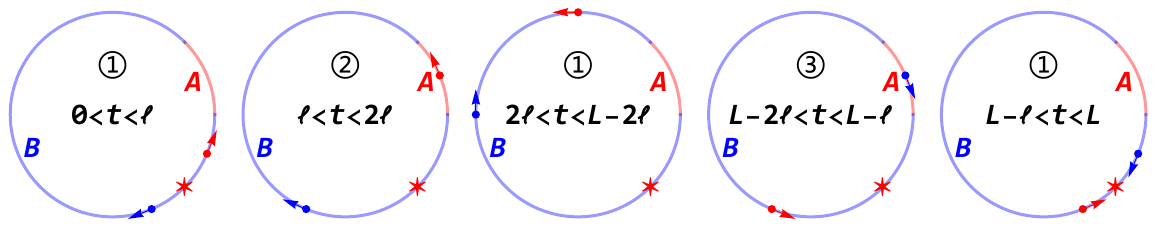}\\
  \caption{Schematic illustration of the quasiparticle picture for the post-quench state after a local operator quench. The subsystem $A=[0,\ell]$ is defined on a spatial circle of total circumference $L$. At $t=0$, a local operator is inserted at position $x=-\ell$, which excites a right-moving quasiparticle (red dot) and a left-moving quasiparticle (blue dot), both propagating at the speed of light $v=1$ in natural units.}
  \label{FigureQuasiparticlePicture}
\end{figure}

We first analyze the local operator quench in the framework of 2D CFT. The time-dependent post-quench state is defined as \cite{Nozaki:2014hna,Nozaki:2014uaa,He:2014mwa}
\be
|\psi(t)\rag = \ep^{-\ii H t} \cO(-\ell) |G\rag.
\ee
In this setup, the system is defined on a spatial circle of total circumference $L$. We prepare the initial state by inserting the local operator $\cO$ at position $x=-\ell$ on the CFT ground state $|G\rag$, and focus our analysis on the finite spatial subsystem $A=[0,\ell]$. A schematic illustration of the quasiparticle picture for the post-quench state is presented in Fig.~\ref{FigureQuasiparticlePicture}.
At $t=0$, the insertion of the local operator $\cO$ excites a pair of counter-propagating quasiparticles: one right-moving and one left-moving, localized at the insertion point $x=-\ell$. These quasiparticles carry entanglement and alter the entanglement structure of the subsystem they are passing through during their propagation. By tracking the real-time positions of the two quasiparticles over one full period $t\in(0, L)$, we resolve the evolution into five distinct time intervals { $(0,\ell),\,(\ell,2\ell),\,(2\ell,L-2\ell),\,(L-2\ell,L-l)$ and $(L-\ell,L)$}, which we categorize into three core dynamical regimes
\bea
&& \text{\ding{192}}~\textrm{no quasiparticle in } A: 0<t<\ell, 2\ell<t<L-2\ell, L-\ell<t<L, \nn\\
&& \text{\ding{193}}~\textrm{right mover in } A: \ell<t<2\ell, \nn\\
&& \text{\ding{194}}~\textrm{left mover in } A: L-2\ell<t<L-\ell.
\eea

Following \cite{Zhang:2019kwu}, we specialize to a class of local operators of the form
\be
\cO(w,\bar w) = \m \cP(w) + \n \cQ(\bar w),
\ee
where $\cP(w)$ is a normalized holomorphic (right-moving) primary operator, and $\cQ(\bar w)$ is a normalized anti-holomorphic (left-moving) primary operator. We define the relative weights of the holomorphic and anti-holomorphic components as
\be
p=\f{|\m^2|}{|\m^2|+|\n^2|}, ~~
q=\f{|\n^2|}{|\m^2|+|\n^2|},
\ee
satisfying $p+q=1$ by construction.

Using the quasiparticle picture, we derive the time-dependent effective RDMs for the two subsystems. For subsystem $A=[0,\ell]$, the RDM takes the piecewise form
\be \label{rAt}
\r_A(t) = \lt\{
\ba{ll}
\r_{A,G}                     & t \in \text{\ding{192}} \\
p \r_{A,\cP}(t) + q \r_{A,G} & t \in \text{\ding{193}} \\
p \r_{A,G} + q \r_{A,\cQ}(t) & t \in \text{\ding{194}}
\ea
\rt.\!\!\!,
\ee
where $\rho_{A,G}$ denotes the RDM of subsystem $A$ in the CFT ground state, $\rho_{A,\mathcal{P}}(t)$ denotes the RDM of subsystem $A$ at time $t$ when a right-moving quasiparticle is present inside subsystem $A$, and $\rho_{A,\mathcal{Q}}(t)$ denotes the RDM of subsystem $A$ at time $t$ when a left-moving quasiparticle is present inside subsystem $A$. For its complementary subsystem $B=[\ell,L]$, the corresponding effective RDM is given by
\be \label{rBt}
\r_B(t) = \lt\{
\ba{ll}
p \r_{B,\cP}(t) + q \r_{B,\cQ}(t) + \sqrt{pq} [ \r_{B,\cP,\cQ}(t) + \r_{B,\cQ,\cP}(t) ] & t \in \text{\ding{192}} \\
p \r_{B,G} + q \r_{B,\cQ}(t) & t \in \text{\ding{193}} \\
p \r_{B,\cP}(t) + q \r_{B,G} & t \in \text{\ding{194}}
\ea
\rt.\!\!\!.
\ee
The following calculations are based on the orthogonality of the effective RDMs. For subsystem $A$, the ground state RDM with no quasiparticle $\rho_{A,G}$ is orthogonal to any single-quasiparticle RDM, say $\rho_{A,\mathcal{P}}(t)$ with $t \in \text{\ding{193}}$, such that $\rho_{A,G}\rho_{A,\mathcal{P}}(t) = 0$. Furthermore, RDMs corresponding to distinct quasiparticles are mutually orthogonal, for instance, $\rho_{A,\mathcal{P}}(t_1) \rho_{A,\mathcal{Q}}(t_2) = 0$ for all $t_1 \in \text{\ding{193}}$ and $t_2 \in \text{\ding{194}}$. In addition, RDMs of the same quasiparticle at different positions are also orthogonal, i.e., $\rho_{A,\mathcal{P}}(t_1) \rho_{A,\mathcal{P}}(t_2) = 0$ for any $t_1, t_2 \in \text{\ding{193}}$ with $t_1 \neq t_2$. Analogous orthogonality conditions apply to subsystem $B$.

\begin{figure}[t]
  \centering
  \includegraphics[width=0.7\textwidth]{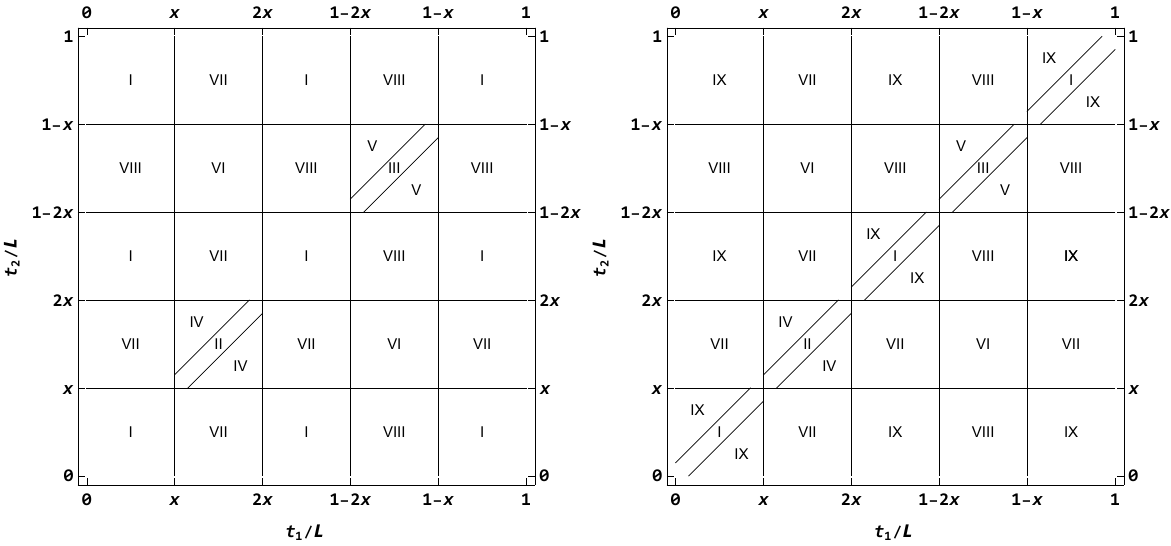}\\
  \caption{Partitioning of the $(t_1,t_2)$ parameter plane for two arbitrary times within one full period $t_1,t_2\in[0,L]$. The left panel shows the eight distinct regions for subsystem $A=[0,\ell]$, while the right panel shows the nine distinct regions for its complementary subsystem $B=[\ell, L]$. The dimensionless subsystem size is defined as $x=\f{\ell}{L}$. Note that the diagonal regimes are actually infinitely thin, i.e., only apply for $t_1=t_2$.}
  \label{FigureDivide}
\end{figure}

To compute the Bures distance and product entropy between two arbitrary times $t_1$ and $t_2$ within a single period $t_1,t_2\in[0, L]$, we partition the $(t_1,t_2)$ parameter plane into distinct regions based on the dynamical regimes of the RDMs. This yields eight distinct regions for subsystem $A$ and nine distinct regions for subsystem $B$, as visualized in Fig.~\ref{FigureDivide}.

To isolate the dynamical contributions to entanglement and correlation measures induced by the quench, we define the excess entanglement entropy and excess subsystem product entropy by subtracting the static ground state contributions
\bea
&& \D S_A(t) \equiv S_A(t) - S^{(1)}(\r_{A,G}), \nn\\
&& \D S_B(t) \equiv S_B(t) - S^{(1)}(\r_{B,G}),
\eea
\bea
&& \D S_A^{(1,m)}(t_1,t_2) \equiv S_A^{(1,m)}(t_1,t_2) - S^{(1,m)}(\r_{A,G},\r_{A,G}), \nn\\
&& \D S_B^{(1,m)}(t_1,t_2) \equiv S_B^{(1,m)}(t_1,t_2) - S^{(1,m)}(\r_{B,G},\r_{B,G}).
\eea

Using the piecewise time-dependent RDMs derived above, Eqs. (\ref{rAt}) and (\ref{rBt}), we analytically obtain the excess entanglement entropy for both subsystems, which takes the following simple form
\be
\D S_A(t) = \D S_B(t) = \lt\{
\ba{ll}
0                     & t \in \text{\ding{192}} \\
- p \log p - q \log q & t \in \text{\ding{193}},\text{\ding{194}}
\ea
\rt.\!\!\!.
\ee

We further compute the subsystem Bures distance between two arbitrary times $t_1$ and $t_2$, with piecewise results for subsystem $A$
\be
D_A(t_1,t_2) = \lt\{
\ba{ll}
0                     & (t_1,t_2) \in \I,\II,\III \\
\sqrt{2(1-q)}         & (t_1,t_2) \in \IV \\
\sqrt{2(1-p)}         & (t_1,t_2) \in \V \\
\sqrt{2(1-\sqrt{pq})} & (t_1,t_2) \in \VI \\
\sqrt{2(1-\sqrt{q})}  & (t_1,t_2) \in \VII \\
\sqrt{2(1-\sqrt{p})}  & (t_1,t_2) \in \VIII
\ea
\rt.\!\!\!,
\ee
and for the complementary subsystem $B$
\be
D_B(t_1,t_2) = \lt\{
\ba{ll}
0                     & (t_1,t_2) \in \I,\II,\III \\
\sqrt{2(1-p)}         & (t_1,t_2) \in \IV \\
\sqrt{2(1-q)}         & (t_1,t_2) \in \V \\
\sqrt{2(1-\sqrt{pq})} & (t_1,t_2) \in \VI \\
\sqrt{2}  & (t_1,t_2) \in \VII,\VIII,\IX \\
\ea
\rt.\!\!\!.
\ee

Finally, we present the excess subsystem product entropy for both subsystems
\be
\D S_A^{(1,m)}(t_1,t_2) = \lt\{
\ba{ll}
0 & (t_1,t_2) \in \I,\IV,\V,\VI,\VII,\VIII \\
- \f{p^m}{p^m+q^m} \log \f{p^m}{p^m+q^m} - \f{q^m}{p^m+q^m} \log \f{q^m}{p^m+q^m} & (t_1,t_2) \in \II,\III
\ea
\rt.\!\!\!,
\ee
\be
\D S_B^{(1,m)}(t_1,t_2) = \lt\{
\ba{ll}
0 & (t_1,t_2) \in \I,\IV,\V,\VI \\
- \f{p^m}{p^m+q^m} \log \f{p^m}{p^m+q^m} - \f{q^m}{p^m+q^m} \log \f{q^m}{p^m+q^m} & (t_1,t_2) \in \II,\III \\
\textrm{not available} & (t_1,t_2) \in \VII,\VIII,\IX
\ea
\rt.\!\!\!.
\ee
We note that for the complementary subsystem $B$, when $(t_1,t_2)$ falls into regions $\VII$, $\VIII$, or $\IX$, the two RDMs at times $t_1$ and $t_2$ are orthogonal in the CFT scaling limit. As a result, the subsystem product entropy cannot be consistently evaluated within the quasiparticle picture for these time regions.

\subsubsection{Non-chiral example}

\begin{figure}[ht]
  \centering
  \includegraphics[width=0.9\textwidth]{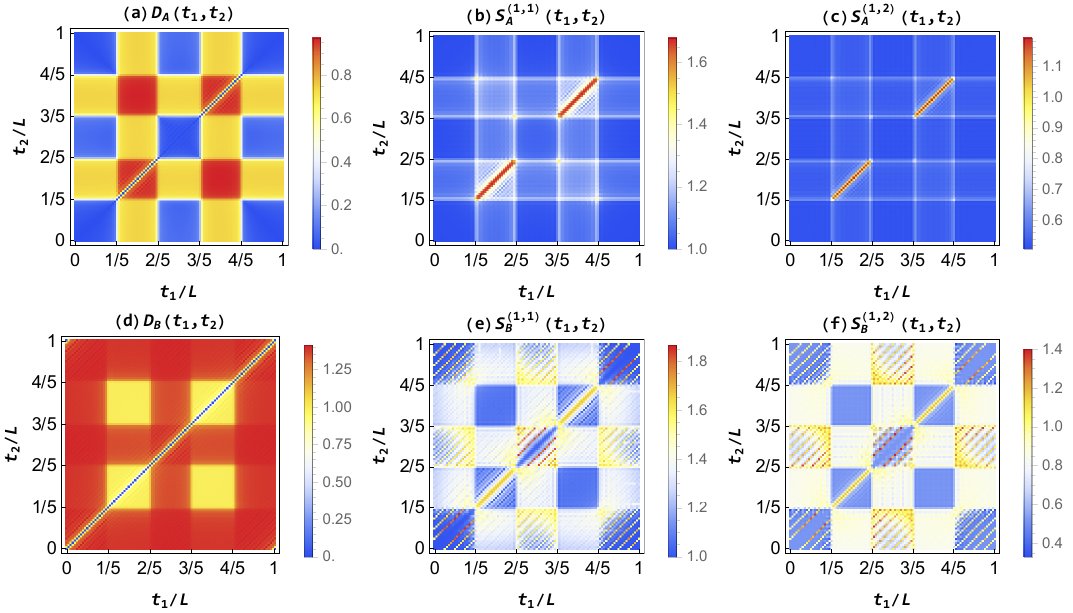}\\
  \caption{Time evolution of the subsystem Bures distance (first column) and subsystem product entropies for $m=1$ (second column) and $m=2$ (third column). Plots are given for subsystems $A=[1,\ell]$ (first row) and $B=[\ell+1,L]$ (second row), following a local operator quench realized by inserting $d_{2j-1}$ at $j=-\ell$ in a critical Ising chain of length $L$. The dispersion relation of the critical Ising chain is linearized throughout the calculation, with parameter values $L=120$ and $\ell=24$.}
  \label{FigureLocalOperatorQuench1}
\end{figure}

\begin{figure}[ht]
  \centering
  \includegraphics[width=\textwidth]{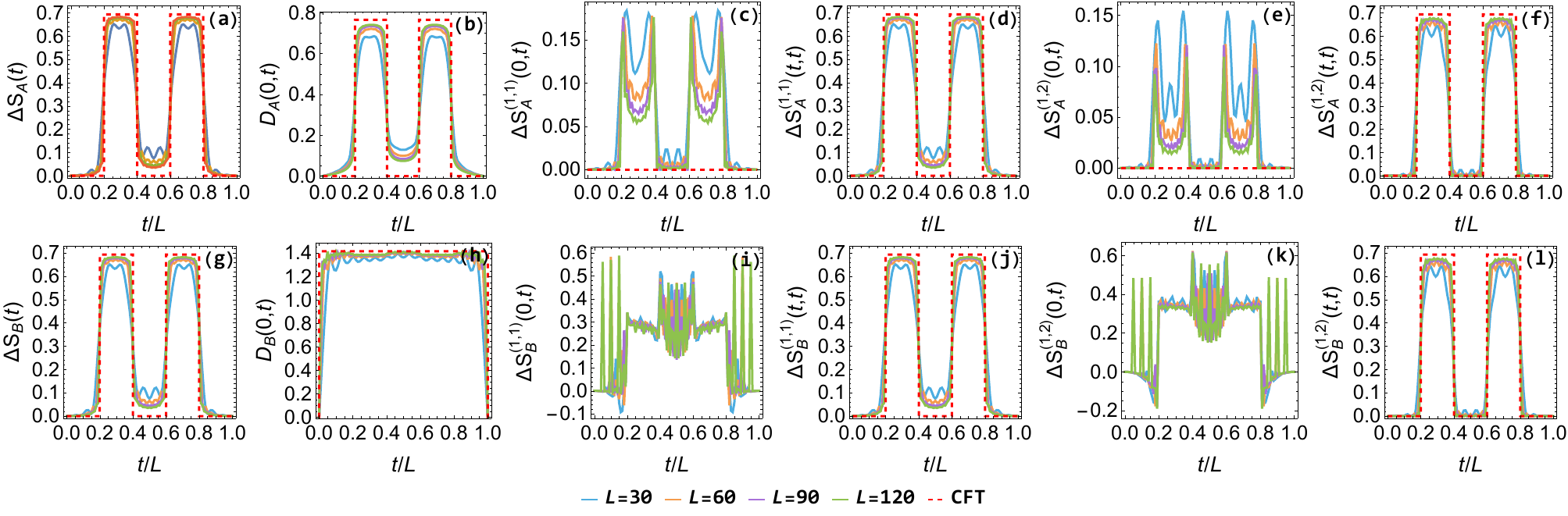}\\
  \caption{Time evolution of the entanglement entropy (first column), subsystem Bures distance (second column), and subsystem product entropies for $m=1$ (third and fourth columns) and $m=2$ (fifth and sixth columns) for subsystems $A=[1,\ell]$ (first row) and $B=[\ell+1,L]$ (second row). The local quench is implemented by inserting $d_{2j-1}$ at $j=-\ell$ in a critical Ising chain of length $L$. Red dashed lines represent the CFT predictions. The deviations in panels (c) and (e) seem to be finite-size effects. Notably, no corresponding CFT results are available for panels (i) and (k) due to nearly orthogonal RDMs in the spin chain picture, which should approach exact orthogonality in the scaling limit.}
  \label{FigureLocalOperatorQuench2}
\end{figure}

We now consider the {operator $d_{2j-1}=(\prod_{i=1}^{L-1}\s_j^z)\s_j^x$} located at $j=-\ell$ within the critical Ising chain at $h=1$.
For this configuration, the weight coefficients of the holomorphic and anti-holomorphic contributions take the non-chiral form \cite{Zhang:2019kwu}
\be
p = q =\f{1}{2}.
\ee
Our numerical results are presented in Figs.~\ref{FigureLocalOperatorQuench1} and \ref{FigureLocalOperatorQuench2}, which show good agreement with the CFT predictions.

\subsubsection{Chiral example}

\begin{figure}[ht]
  \centering
  \includegraphics[width=0.9\textwidth]{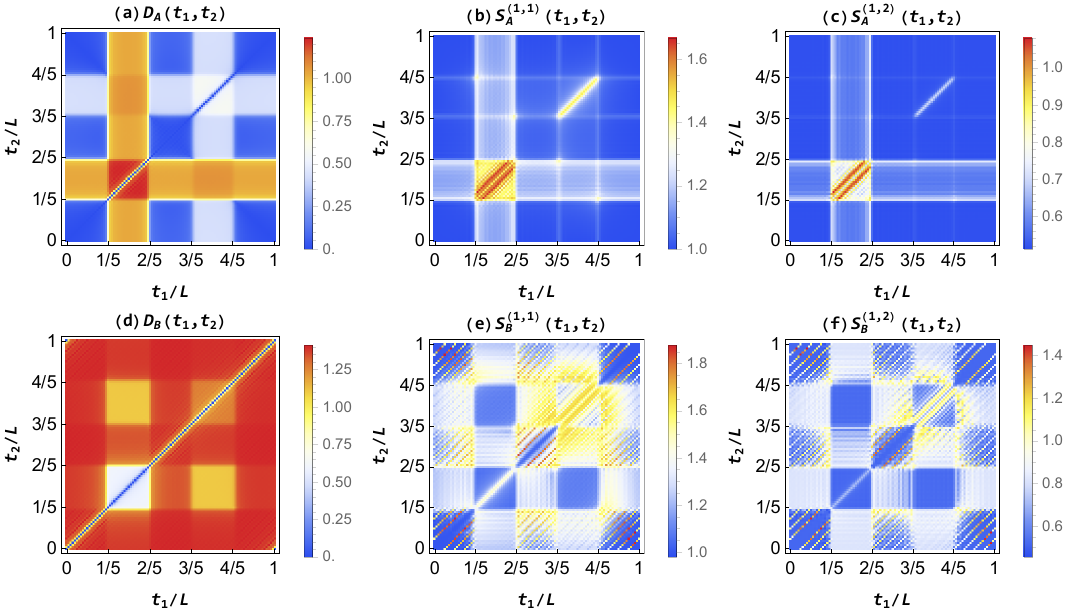}\\
  \caption{Time evolution of the subsystem Bures distance (first column) and subsystem product entropies for $m=1$ (second column) and $m=2$ (third column) for subsystems $A=[1,\ell]$ (first row) and $B=[\ell+1, L]$ (second row). The local operator quench is implemented by inserting $d_{2j-1}+d_{2j}$ at $j=-\ell$ in a critical Ising chain of length $L$. We adopt a linearized dispersion relation for the critical Ising chain and set $L=120$ and $\ell=24$.}
  \label{FigureLocalOperatorQuench3}
\end{figure}

\begin{figure}[ht]
  \centering
  \includegraphics[width=\textwidth]{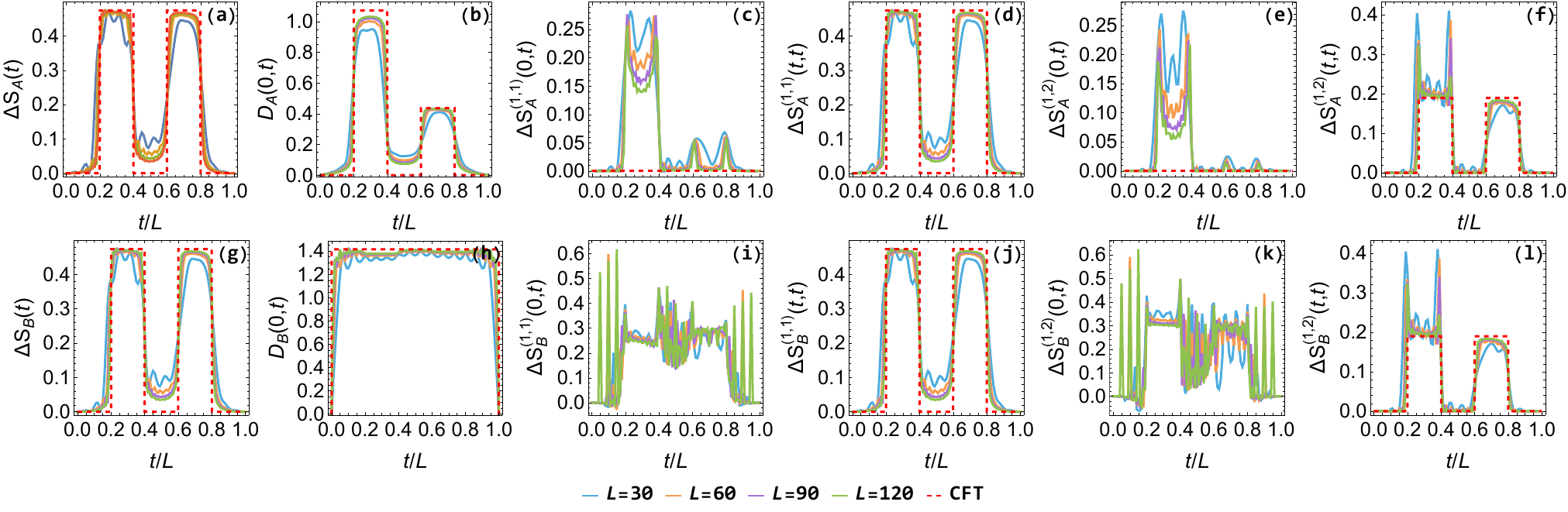}\\
  \caption{Time evolution of the entanglement entropy (first column), subsystem Bures distance (second column), and subsystem product entropies for $m=1$ (third and fourth columns) and $m=2$ (fifth and sixth columns) for subsystems $A=[1,\ell]$ (first row) and $B=[\ell+1, L]$ (second row). The local operator quench is performed by inserting $d_{2j-1}+d_{2j}$ at $j=-\ell$ in a critical Ising chain of length $L$. Red dashed lines correspond to the CFT predictions.  Note that the deviations from the CFT result in (c) and (e) are due to finite-size effects, and no CFT results are available for panels (i) and (k) due to near orthogonality of their RDMs.}
  \label{FigureLocalOperatorQuench4}
\end{figure}

We next examine the composite operator $d_{2j-1}+d_{2j}$ positioned at $j=-\ell$ in the critical Ising chain with $h=1$. {These two operators are defined as $d_{2j-1}=(\prod_{i=1}^{L-1}\s_j^z)\s_j^x$ and $d_{2j}=(\prod_{i=1}^{L-1}\s_j^z)\s_j^y$, respectively.} For this chiral setup, the relative weights of the holomorphic and anti-holomorphic components read \cite{Zhang:2019kwu}
\be
p = \f{1}{2} + \f{1}{\pi}, ~~
q = \f{1}{2} - \f{1}{\pi}.
\ee
Numerical results are presented in Figs.~\ref{FigureLocalOperatorQuench3} and \ref{FigureLocalOperatorQuench4}, and again show good agreement with the CFT predictions. { We expect that the spin chain approaches the CFT results in the scaling limit. Panels (c) and (e) seem to defy this at first glance, but as the values of product entropies in these plots are getting smaller and smaller for larger system size, we believe that they will approach zero in the scaling limit, and that the deviations from the CFT result shown in these panels are due to finite-size effects.}

\section{Conclusion and Outlook} \label{sectionCon}

In this work, we introduce the subsystem product entropy as a novel quantity that unifies and generalizes a broad class of existing entanglement measures and state distinguishability quantifiers. Defined as the R\'enyi (or von Neumann) entropy of a normalized product operator constructed from a pair of non-orthogonal density matrices, this entropy admits a duality: the SVD entanglement entropy of subsystem $A$ for two pure states is equivalent to the subsystem product entropy of its complementary subsystem $B$. This equivalence not only provides a computationally efficient scheme that circumvents direct manipulation of the non-hermitian reduced transition matrix, but also endows the SVD entanglement entropy with a physical interpretation in terms of the spectral diversity of the product between two subsystem states.

In essence, the subsystem product entropy bridges the gap between the entanglement structure of individual quantum states and subsystem overlap measures between pairs of states. It quantifies the detailed structure of the overlap between two quantum states by resolving, at the subsystem level, the spectral distribution of their product. This characterization is insensitive to the magnitude of the overlap, provided the overlap is nonvanishing. Combining the merits of universality, computational efficiency, and experimental accessibility, this quantity opens possible new connections between quantum information science, condensed-matter many-body physics, conformal field theory, and holographic quantum gravity.

We calculate the product entropy in three representative physical settings, namely low-lying eigenstates in 2D CFTs and nonequilibrium dynamics under both global and local quantum quenches, deriving corresponding closed-form analytical results where possible. All analytical predictions are verified numerically, where possible in the critical transverse-field Ising chain, with excellent agreement observed between numerical computations and CFT results up to finite-size corrections.

Our results open up several directions for future research. First, as a unified theoretical framework encompassing various recently proposed quantum information measures, the subsystem product entropy merits further exploration of its operational interpretations in contexts such as quantum hypothesis testing and quantum resource theories. In particular, an analysis of its distillation and monotonicity properties following the investigation of SVD and ABB entropies in \cite{Chen:2025ibe} would be very useful. In addition, the duality between the SVD entanglement entropy of a subsystem and the product entropy of its complement can be extended to holographic systems. A direct geometric dual formulation of the subsystem product entropy may uncover novel connections between quantum information theory and quantum gravitational physics, in analogy with holographic studies of pseudo-entropy \cite{Nakata:2020luh} and SVD entanglement entropy \cite{Parzygnat:2023avh}.

The present CFT analysis is limited to free massless fermions and simple operator excitations. The replica formalism developed in this work can be generalized to arbitrary CFTs, descendant states, and multi-excitation configurations. Evaluating the subsystem product entropy for thermal states, as well as systems with boundaries and defects, may reveal universal topological signatures. Furthermore, investigating its properties under symmetry resolution along the lines of \cite{goldstein2018symmetry,zhao2021symmetry,di2023boundary,northe2023entanglement} might reveal even fine-grained (topological) information about the quantum systems in question. In higher-dimensional or massive field theories, one expects qualitatively distinct scaling behaviors, such as area-law entanglement.

The quasiparticle picture accurately captures the dynamical evolution of these information-theoretic quantities. Extending these analyses to interacting integrable models, which host richer quasiparticle spectra and nontrivial scattering processes, may reveal novel dynamical regimes and corrections beyond the standard semiclassical quasiparticle description. Notably, the plateaus emerging in the subsystem product entropy during local operator quenches directly encode the chiral content of the underlying elementary excitations, making the subsystem product entropy a promising observable for experimental characterization in quantum simulators, especially in platforms where RDMs are accessible. A particularly compelling direction is the experimental implementation of the measurement protocol discussed in Subsection~\ref{measurement}.

\section*{Acknowledgements}

This research was partly supported by NSFC Grant Nos. 11735001, 12275004, 12475053, 12235016, 12588101, and 12205217.
R.M.~acknowledges the support of the German Research Foundation (DFG) through the Collaborative Research Center ToCoTronics, Project-ID 258499086 - SFB 1170, as well as Germany's Excellence Strategy through the W{\"u}rzburg-Dresden Cluster of Excellence on Complexity and Topology in Quantum Matter - ctd.qmat (EXC 2147, Project-ID 390858490).
R.M.~furthermore acknowledges hospitality from the Shanghai Institute for Mathematics and Interdisciplinary Sciences (SIMIS)  and associated travel support under STCSM Grant 25HB2701900.
Z.-Y.X.~acknowledges support from the Berlin Quantum Initiative.
J.Z.~is also supported by the Tianjin University Self-Innovation Fund Extreme Basic Research Project (Grant No.~2025XJ21-0007).
Numerical calculations for this study were performed at a high-performance cluster at the Center for Joint Quantum Studies (HPC-CJQS) of Tianjin University.

%the National Natural Science Foundation of China (Grant No.~12205217) and

%\appendix

%\bibliographystyle{D:/00.bibx/JHEPx}
%\bibliography{D:/00.bibx/2026,D:/00.bibx/2025,D:/00.bibx/2024,D:/00.bibx/2023,D:/00.bibx/2022,D:/00.bibx/2021,D:/00.bibx/2020,D:/00.bibx/2019,D:/00.bibx/2018,D:/00.bibx/1960,D:/00.bibx/1970,D:/00.bibx/1980,D:/00.bibx/1990,D:/00.bibx/1995,D:/00.bibx/1996,D:/00.bibx/1997,D:/00.bibx/1998,D:/00.bibx/1999,D:/00.bibx/2000,D:/00.bibx/2001,D:/00.bibx/2002,D:/00.bibx/2003,D:/00.bibx/2004,D:/00.bibx/2005,D:/00.bibx/2006,D:/00.bibx/2007,D:/00.bibx/2008,D:/00.bibx/2009,D:/00.bibx/2010,D:/00.bibx/2011,D:/00.bibx/2012,D:/00.bibx/2013,D:/00.bibx/2014,D:/00.bibx/2015,D:/00.bibx/2016,D:/00.bibx/2017,D:/00.bibx/book,D:/00.bibx/work,D:/00.bibx/thesis}

%\bibliographystyle{unsrt}
%\bibliography{SubsystemProducEntropy_checked_v260910}

%\iffalse

%\fi

\end{document}